\documentstyle [11pt,epsf] {article}

\catcode`@=12
\newcommand{\be}{\begin{equation}}
\newcommand{\ee}{\end{equation}}
\newcommand{\ber}{\begin{eqnarray}}
\newcommand{\eer}{\end{eqnarray}}
\newcommand{\bers}{\begin{eqnarray*}}
\newcommand{\eers}{\end{eqnarray*}}

\begin{document}
\vspace{0.5in}
\oddsidemargin -.375in  
\newcount\sectionnumber 
\sectionnumber=0 

\def\be{\begin{equation}} 
\def\ee{\end{equation}}
\thispagestyle{empty}
\begin{flushright} UH-511-930-99 \\
UT-PT-99-08\\ May 1999\\
\end{flushright}
\vspace {.5in}
\begin{center}
{\Large \bf{Measuring $\beta$ in $B \to D^{(*)+}D^{(*)-}K_s$ 
Decays\\}}
\vspace{.5in}
{\rm {T.E. Browder$^1$, A. Datta$^2$,  
 P. J. O'Donnell$^2$ and S. Pakvasa$^1$} \\}

\vskip .3in

{\it $^1$Department of Physics and Astronomy \\}
{\it  University of Hawaii, \\}
{\it Honolulu, Hawaii 96822\\}

{\it and\\}

{\it $^2$Department of Physics  \\}
{\it  University of Toronto, \\}
{\it Toronto, Ontario, Canada M5S 1A7\\}

%\vspace{.1in}
\vskip .5in
\end{center}  
\begin{abstract}
We consider the possibility of measuring both $\sin (2 \beta)$ and
$\cos ( 2 \beta)$ in the KM
unitarity triangle using the process $B^0 \to D^{*+}D^{*-}K_s$.
This decay mode has a higher branching fraction (O(1\%)) than 
the mode $B^0 \to D^{*+}D^{*-}$. 
We use the factorization assumption and heavy hadron chiral
perturbation theory to estimate the branching fraction and polarization.
The time dependent rate for
$B^0(t) \to D^{*+} D^{*-} K_s$ 
 can be used to measure $\sin (2 \beta)$ and $\cos(2 \beta)$ . Furthermore, 
examination of the $D^{*+} K_s$ mass spectrum
may be the best way to experimentally find the broad $1^+$ p-wave
$D_s$ meson. 

 \end {abstract}
\newpage
\baselineskip 24pt

%
% Turn this off to get double spaced version
%\tighten
%
%

\section{Introduction}
The decay $B^0 \to J/\psi K_s$ is expected to
provide a clean measurement of the angle $\sin (2\beta)$ in
the unitarity triangle \cite{Sanda}. However, other
modes can also provide relevant information on the angle $\beta$.
An example of such a mode is the decay $B^0 \to D^{(*)} {\overline D}^{(*)}$.
In this mode $ B^0 \to D^{*+} {D}^{*-}$, the vector-vector final state
in general is an admixture of CP odd and even eigenstates, 
 because s, p and 
d partial waves with different CP-parities can contribute. Since the CP
 asymmetry
has opposite sign for the two CP states they tend to cancel or dilute the 
overall asymmetry. The amount of dilution of the CP asymmetry is 
represented by the dilution factor, $D$, which depends on the CP 
composition of the final state. 
 The presence of two CP components in 
the final state of $ B^0 \to D^{*+} {D}^{*-}$ makes the dilution 
factor, $D < 1$ for this decay. This is
unlike the
case for a mode such as $B^0 \to D^{+}D^{-}$ where the final state is
a CP eigenstate and D=1 as there is no dilution of the CP asymmetry. An 
angular analysis can  extract the contribution of the 
different CP eigenstates leading to a measurement of $D$ and hence
 of $\sin(2 \beta)$ 
\cite{Dunietz, Michelon}. However, in the factorization approximation 
and using Heavy 
Quark Effective Theory (HQET) it can be shown that the final state in 
$ B^0 \to D^{*+} { D}^{*-}$ is dominated by a single CP 
eigenstate \cite{Rosner}. To the extent that
this is valid, the angle $\sin (2 \beta)$ can be 
determined without the need for an angular analysis.  The decay
$ B^0 \to D^{*+} { D}^{*-}$ may be preferred to 
$ B^0 \to D^{+} { D}^{-}$ because contamination from 
penguin contributions and final state interactions (FSI) are
expected to be smaller in the former decay \cite {Michelon}.

In this work we consider the possibility of extracting $\beta$ from the decay
$ B^0 \to D^{(*)} {\bar D}^{(*)} K_s$. These modes are enhanced relative to
$ B^0 \to D^{(*)} {\overline D}^{(*)}$  by the 
factor $|V_{cs}/V_{cd}|^2 \sim 20$.
As in the case of $B^0 \to J/\psi K_s$
decay, the penguin contamination is expected to be small in these decays. 
Moreover these decays can be used to probe both $\sin 2 \beta$ and 
$ \cos {2 \beta}$ which can resolve
$\beta \rightarrow \pi/2- \beta$ ambiguity \cite{Charles}.

The possibility that a large portion of the $b\to c \bar{c} s$
decays materialize as $B\to D\bar{D} K$ was first suggested
by Buchalla {\it et al.}\cite{bdy}. Using wrong sign $D$-lepton
correlations, experimental evidence for this possibility was
found by CLEO, who observed ${\cal B}(B\to D X)= (7.9\pm 2.2)\%$\cite{dlep}. 
Later, CLEO\cite{cleoddk}, 
ALEPH\cite{alephddk} and DELPHI\cite{delphiddk} reported full reconstruction of
exclusive $D \bar{D} K$ final states with branching fractions that
are consistent with the result from $D$-lepton correlations.

CLEO obtained 
${\cal B}({\overline {B}}^0\to D^{*+} \bar{D}^{*0} K^-) = 
(1.30^{+0.61}_{-0.47}\pm 0.27)\%$ and
${\cal B}(B^-\to D^{*0} \bar{D}^{*0} K^-) = 
(1.45^{+0.78}_{-0.58} \pm 0.36)\%$. These  values should be 
approximately equal to the branching fraction for 
${\cal B}(B^0\to D^{*+} D^{*-} K^0)$. We use the latter value
for the purpose of a sensitivity estimate. Taking into account
${\cal B}(K^0\to K_s)=0.5$, ${\cal B}(K_s\to \pi^+ \pi^-)=0.667$,
and assuming that the $K_s$ reconstruction efficiency is $\sim 0.5$, we
can estimate the ratio of the tagged $B^0\to D^{*+} D^{*-} K_s$ 
events to the tagged $D^{*+} D^{*-}$ events. Assuming ${\cal
B}(B^0\to D^{*+} D^{*-})=6\times 10^{-4}$, which is the central value
of the recent CLEO measurement\cite{cleodd}, we find that the ratio of
the number of events is $\sim 4.0$. Therefore, this mode could be more
sensitive to the CP violation angle $\sin(2\beta)$ 
than $B^0\to D^{*+} D^{*-}$. However, if the final state contains a resonance
then $B^0$ and ${\bar{B}}^0$ can be distinguished and there
is additional dilution of the CP asymmetry. For the decay $B \to f$ and
$ {\bar B} \to { \bar f}$ the dilution factor,
 $D$, measures the ratio of
 the overlap of the amplitudes for $B \to f$ and ${\bar B} \to { \bar f}$ 
  to the average of the  decay rate for $B \to f$ and ${\bar B} \to {\bar f}$ .
Clearly $D=1$ when the amplitudes for
$B \to f$ and ${\bar B} \to { \bar f}$ decays are equal. When the final state
in the decay $ B \to D^{*+} D^{*-} K_s$  
 contains a resonance the amplitude for 
$B$ and ${\bar B}$ decays are different because the resonance 
in the $B$ and ${\bar B}$ final state
   occur
at different kinematical points. This causes additional mismatch of
the 
$B$ and ${\bar B}$ amplitudes  which results in the further dilution of 
the CP asymmetry. 
  A similar
conclusion is obtained in the comparison of $B^0\to D^+ D^- K_s$  
to $B^0\to D^+ D^-$. The above conclusions are detector
dependent; a somewhat pessimistic estimate of the
$K_s$ reconstruction efficiency is used here while the detection
efficiency for the $D^{*+} D^{*-}$ final state is assumed to be
similar for both cases. Better determination
of the CP sensitivities will require more
precise measurements of the branching fractions for
the $D^{*} \bar{D}^{*} K$ decay modes and will also depend
on details of the experimental apparatus and reconstruction programs.

The amplitude for the decay $B^0\to D^* \bar{D}^* K_s$ can  
have a resonant contribution 
and a non-resonant contribution.
 For the resonant contribution
 the $D^* K_s$ in the final state comes dominantly from an
excited $D_s (1^+)$ state.
In the approximation of treating $D^* \bar{D}^* K_s$ as $D^* D_s(excited)$,
there are four possible excited p-wave $D_s$ states which might contribute.
These are the two  states with the light degrees of freedom
in a $j^P=3/2^{+}$ state and the two  states with light
degrees of freedom in a $j^P = 1/2^{+}$ state.
Since the states with $j^P=3/2^{+}$ decay via d-wave to $D^* K_s$, 
they are suppressed. Of the states
with light degrees of freedom in $j^P=1/2^{+}$
states, only the $1^+$ state contributes. The $0^+$ state is forbidden
to decay to the final state $D^* K_s$.

To estimate the above contribution and to calculate the non-resonant
amplitude, we use heavy hadron chiral perturbation theory 
 (HHCHPT)\cite{HHCHPT}.  The momentum $p_k$ of $K_s$ can have a 
maximum value of about  1 GeV for 
$ B^0 \to D^{*+} {\bar D}^{*-} K_s$. This is of the same order as 
$\Lambda_{\chi}$ which sets the scale below which  
we expect HHCHPT to be valid. It follows that in the present case it is 
reasonable to apply HHCHPT to calculate the three body decays. 

In the lowest order in the
HHCHPT expansion, contributions to the decay amplitude come from the 
contact interaction terms and the pole diagrams which give rise to the 
non-resonant and resonant contributions respectively. The pole diagrams get 
contributions from the various 
multiplets involving $D_s$ type resonances as mentioned above. 
In the framework of HHCHPT, the
 ground state heavy meson has the light degrees of freedom in a
spin-parity state $j^P={1\over2}^-$, corresponding to the usual
pseudoscalar-vector meson doublet with $J^P=(0^-,1^-)$.  The first
excited state involves a p-wave excitation, in which the light
degrees of freedom have $j^P={1\over2}^+$ or ${3\over2}^+$.  In the
latter case we have a heavy doublet with $J^P=(1^+,2^+)$. These states can
probably be identified with $D_{s1}(2536)$ and $D_{sJ}(2573)$ \cite{PDG}.
Heavy quark symmetry rules out any pseudoscalar coupling of this
doublet to the ground state at lowest order in 
the chiral expansion \cite{Fluke};
hence the effects of these states will be suppressed and
we will ignore them in our analysis.  
In fact there is  an experimental upper limit on inclusive
$B \to D_{s1}(2536) X<0.95 \%$ at 90\% C.L \cite{cleoD}. Since the total 
$D^*{\overline D}^* K$
rate is about 8\%, this confirms that the narrow p-wave states
do not account for a significant fraction of the total $D^* \bar{D}^* K$ rate.

The other excited doublet has
$J^P=(0^+,1^+)$.  
These states are expected to decay rapidly through
s-wave pion emission and have large widths \cite{nathanmark}. Observation of 
the $1^+$ state in the $D$ system was recently reported by CLEO \cite{cleores}.
Only the $1^+$ can contribute in this case. For later reference,
we denote this state by $D^{*'}_{s1}$.  
However, quark model estimates suggest \cite{God}
that these states
should have masses near $m+\delta m$ with $\delta m=500$ MeV, 
where $m$ is the mass of the lowest multiplet.

We will assume that the leading order terms in HHCHPT give 
the dominant contribution to the decay amplitude 
and so we will neglect all sub-leading effects 
suppressed by $1/\Lambda_{\chi}$ and
  $1/m$, where m is the heavy quark mass. We  show that from the 
time dependent analysis of 
 $ B^0(t) \to D^{*+} { D}^{*-}K_s$ one can extract
 $\sin (2\beta)$ and $\cos (2\beta)$.
 Measurement of both
$\sin (2\beta)$ and $\cos (2\beta)$ can resolve the
$\beta \rightarrow \pi/2- \beta$ ambiguity \cite{Charles,Gros,Dig}.  
The measurement of
 $\sin (2\beta)$ can be made from the time dependent partial rate asymmetry
while a fit to the time dependent rate for
$\Gamma[ B^0(t) \to D^{*+} { D}^{*-}K_s] +
\Gamma[ {\bar B}^0(t) \to D^{*+} { D}^{*-}K_s] $ may be used for
the extraction of $\cos (2\beta)$. Note that the
$\cos (2\beta)$ term measures  the overlap of the imaginary part of
the amplitudes for $B \to D^{*+} D^{*-} K_s$ and 
${\bar B} \to D^{*+}D^{*-}K_s$ decays and is non zero only if there is a 
resonance contribution.

As in the case for $ B \to D^{*+} { D}^{*-}$ the asymmetry in
$B \to D^{*+} { D}^{*-} K_s$ is also diluted.
For the non resonant contribution to $B \to D^{*+} { D}^{*-} K_s$ 
 the final state is an admixture of CP states 
with different CP parities.
This leads to the dilution of the asymmetry and this is the same 
dilution of the asymmetry 
as in the case for
$B \to D^{*+} D^{*-}$.
As already mentioned above, when the resonant contribution is included there 
is further dilution of the asymmetry from the additional mismatch of
the amplitudes for
 $B$ and ${ \bar B}$ decays. One can reduce the additional dilution
of the CP asymmetry by imposing cuts to remove the resonance. A narrow 
resonance is preferable as it can be more effectively removed 
from the signal region than a broad
resonance . In this work
we examine several cuts that can be used to remove the resonance
and lessen the dilution of the CP asymmetry.
When we include  the resonance contribution  we find that
a broader resonance leads to a larger value of D and is a more
 useful  probe of $\cos (2\beta)$  because
of the 
 the larger overlap of the amplitudes for
$B \to D^{*+} { D}^{*-} K_s$  and 
${\bar B} \to D^{*+} { D}^{*-} K_s$ decays.

We also point out that
from the differential decay distribution of the time independent process
$ B^0 \to D^{*+} { D}^{*-}K_s$  one can discover the 
$1^{+}$ resonance $D^{*'}_{s1}$.
We show that the differential decay distribution for  
 small values of $E_k$, the kaon energy, 
shows a clear resonant structure which comes from the pole 
contribution to the amplitude  with the excited $J^P=1^+$
 intermediate state. Therefore, 
examination of the $D^{*} K_s$ mass spectrum
may be the best experimental way to find the broad $1^+$ p-wave
$D_s$ meson and  a fit to the
decay distribution will measure its mass and the coupling.

A similar analysis can be performed for
$ B^0 \to D^{+} { D}^{-} K_s$ \cite{Charles,Narduli}. However the 
predictions of HHCHPT for this mode may be less reliable 
because of the larger energy of the $K_s$. The effects of 
penguin contributions, though small, may also be more important in
$ B^0 \to D^{+} { D}^{-} K_s$ than in
$ B^0 \to D^{*} { D}^{*} K_s$ as in the two body case 
\cite{Michelon}. 

In the next section we describe the extraction of
of $\sin {2 \beta}$ and
$\cos{2\beta}$ from the time dependent rate for
$B(t) \to D^{*+}D^{*-} K_s$. In the next section we present the
 the amplitude for $ B \to D^{*} {\bar D}^{*} K_s$ in the 
factorization approximation and using HHCHPT. 
In the last section we discuss and present our results.

\section{ Extraction of $\sin {2 \beta}$ and $\cos{2\beta}$ } 
In this section we discuss the  extraction of $\sin {2 \beta}$ and
$\cos{2\beta}$ from the time dependent rate for
$B(t) \to D^{*+}D^{*-} K_s$. We define the following amplitudes
\be
{a^{\lambda_1,\lambda_2}} \equiv
 A(B^0(p)\to D^{+*}_{\lambda_1}(p_{+}) D^{-*}_{\lambda_2}(p_{-}) K_s(p_k)),\qquad 
{\bar a^{\lambda_1,\lambda_2}} \equiv 
A(\bar B^0(p)\to D^{+*}_{\lambda_1}(p_{+}) D^{-*}_{\lambda_2}(p_{-}) K_s(p_k  ),\label{AAbar}
\ee
where $B^0$ and $\bar  B^0$ represent unmixed neutral $B$ and $\lambda_1$ and $\lambda_2$ are the polarization indices of the $D^{*+}$ and $D^{*-}$ respectively.

The time-dependent amplitudes for an oscillating state $B^0(t)$ which has been tagged as a $B^0$ meson at time $t=0$ is given by, 
\ber
A^{\lambda_1,\lambda_2}(t)& =& {a}^{\lambda_1,\lambda_2} \cos\left(\frac{\Delta m\,t}2\right) + i e^{-2i \beta}
{\bar{a}}^{\lambda_1,\lambda_2} \sin\left(\frac{\Delta m\,t}2\right)\,,
\eer
and the time-dependent amplitude squared summed over polarizations and integrated over the phase space angles is:
\ber
|A(s^+,s^-;t)|^2 & = & \frac{1}{2}\left[ {\rm G}_0(s^+,s^-)+{\rm G}_{\rm c}(s^+,s^-)\cos\Delta m\,t-{\rm G}_{\rm s}(s^+,s^-)\sin\Delta m\,t \right] \, 
\label{osc}
\eer
with
\begin{eqnarray}
{\rm G}_0(s^+,s^-)     & = & |{a}(s^+,s^-)|^2 +|\bar{{a}}(s^+,s^-)|^2 ,\\
{\rm G}_{\rm c}(s^+,s^-) & = & |{a}(s^+,s^-)|^2 -|\bar{{a}}(s^+,s^-)|^2 ,\\
{\rm G}_{\rm s}(s^+,s^-) & = & 2\Im\left (e^{-2i \beta} \bar{a}(s^+,s^-)
{{a}^\ast(s^+,s^-)} \right ) \nonumber \\
& = & -2 \sin(2\beta)\, \Re \left ( \bar{a}{{a}^\ast} \right ) + 2\cos (2\beta) \,\Im \left ( \bar{a}{{a}^\ast} \right ).\label{gmix}
\end{eqnarray}
The variables $s^+$ and $s^-$ are the Dalitz plot variable
$$ s^+= (p_{+} +p_k)^2,\quad s^-= (p_{-} +p_k)^2$$
The transformation defining the CP-conjugate channel 
$\bar B^0(t)\to D^{*-}D^{*+} K_s$ is $s^+\leftrightarrow s^-$, 
${ a}\leftrightarrow \bar{ a}$ and $\beta\to -\beta$. Then:
\ber
|\bar A(s^-,s^+;t)|^2 & = &\frac{1}{2}\left[ {\rm G}_0(s^-,s^+)-{\rm G}_{\rm c}(s^-,s^+)\cos\Delta m\,t+{\rm G}_{\rm s}(s^-,s^+)\sin\Delta m\,t \right]\,.
\eer
Note that for simplicity the $e^{-\Gamma t}$ and constant phase space factors have
been omitted in the above equations.

It is convenient in our case to replace the variables $s^+$ and $s^-$ by 
the variables $y$ and $E_k$ where $E_k$ is the $K_s$ energy in the rest 
frame of the $B$ and $y=\cos{\theta}$ with $\theta$ being the angle between 
the momentum of $K_s$ and $D^{*+}$ in a frame where the two $ D^{*}$ are moving back to back along the z- axis. This frame is boosted with respect to the rest frame of the $B$ with ${\vec {\beta}}=-({\vec{p}_k}/m_B)(1/(1-E_k/m_B))$. Note
$s^+\leftrightarrow s^-$ corresponds to $y \leftrightarrow -y$. The 
variable $y$ can be expressed in terms of variables in the rest frame of $B$. 
For instance
\bers
E_{+} & = & \frac{E_B^{\prime} E_{+}^{\prime} -{p}_B^{\prime} 
{p}_{+}^{\prime}y}{m_B} \
\eers
where  $E_{+}$ and $E_{+}^{\prime}$ are the energy of the $D^{*+}$ in 
the rest frame of the $B$ and in the boosted frame while $ E_B^{\prime}$ 
is the 
energy of the $B$ in the boosted frame. The magnitudes of the 
momentum of the $B$ and the $D^{*+}$ in the boosted frame are 
given by ${p}_B^{\prime}$ and ${p}_{+}^{\prime}$ respectively.

In the approximation of neglecting the penguin contributions to the 
amplitude there is no direct CP violation. This leads to the relation
\be
{ a}^{\lambda_1,\lambda_2}(\vec{p}_{k1},E_k)=
{\bar a}^{-\lambda_1,-\lambda_2}(\vec{-p}_{k1},E_k)\
\ee
where $\vec{p}_{k1}$ is the momentum of the of the $K_s$ in the boosted frame.
The above relations then leads to 
\ber
G_0(-y,E_k) & = & G_0(y,E_k) \\
G_c(-y,E_k) & = & - G_c(y,E_k) \\
G_{s1}(-y,E_k) & = & G_{s1}(y,E_k) \\
G_{s2}(-y,E_k) & = & -G_{s2}(y,E_k) \
\eer
 where we have defined
\ber
G_{s1}(y,E_k) & = &\Re \left ( \bar{a}{{a}^\ast} \right)   \\
G_{s2}(-y,E_k) & = & \Im \left ( \bar{a}{{a}^\ast} \right)  \
\eer
 Carrying out the integration over the phase space variables $y$ and $E_k$ one 
gets the following expressions for the time-dependent total rates
for $ B^0(t) \to D^{*+}D^{*-}K_s$ and the CP conjugate process
\ber
\Gamma(t) &= &\frac{1}{2}[I_0 + 2 \sin( 2\beta)\sin (\Delta m t) I_{s1}]\\
{\overline{\Gamma}}(t) &= &\frac{1}{2}[I_0 - 2 \sin( 2\beta)
\sin (\Delta m t) I_{s1}]\
\eer
where $I_0$ and $I_{s1}$ are the integrated $G_0(y,E_k)$ and 
$G_{s1}(y,E_k)$ functions.
One can then extract $\sin (2 \beta)$ from the rate asymmetry
\be
\frac{\Gamma(t)-{\overline{\Gamma}}(t)}{\Gamma(t)+{\overline{\Gamma}}(t)} =
D\sin( 2\beta)\sin (\Delta m t)\
\ee
where
\be 
 D = \frac{2I_{s1}}{I_0} \
\ee
is the dilution factor.

The $\cos(2 \beta)$ term can be probed by by integrating over half the 
range of the variable $y$ which can be taken for instance to be $y \ge 0$.
In this case we have
\ber
{\Gamma}(t) &= &\frac{1}{2}[{J}_0 
+{J}_c \cos (\Delta m t) + 2 \sin( 2\beta)\sin (\Delta m t) {J}_{s1}-
2 \cos( 2\beta)\sin (\Delta m t) {J}_{s2}]\\
{\overline{\Gamma}}(t) &= & \frac{1}{2}[{J}_0 
+{J}_c \cos (\Delta m t) - 2 \sin( 2\beta)\sin (\Delta m t) {J}_{s1}-
2 \cos( 2\beta)\sin (\Delta m t) {J}_{s2}]\
\eer 
where ${J}_0$, ${J}_c$, ${J}_{s1}$ and  ${J}_{s2}$,  
are the integrated $G_0(y,E_k)$, $G_c(y,E_k)$, $G_{s1}(y,E_k)$ and 
$G_{s2}(y,E_k)$ functions integrated over the range $y \ge 0$.
One can measure $\cos(2 \beta)$  by fitting to the time distribution
of $\Gamma(t) +{ \bar \Gamma}(t)$.
Measurement of the  $\cos(2 \beta)$ can resolve    
the $\beta \to {\pi \over 2} - \beta$ ambiguity.

\section{Amplitude and Decay Distribution }

 In this section we present the amplitude and decay distribution for
the decay $B \to D^{*+} D^{*-}K_s$ . Details of the calculation 
of the amplitudes using the factorization assumption and HHCHPT
 are given in Appendix A.

 The non-resonant amplitude for the three body decay 
${\bar B}^0(v,m) \to D^{*+}(\epsilon_1,v_{+},m_1) 
D^{*-}(\epsilon_2,v_{-},m_2) K_s(p_k)$, after setting $m_2=m_1$, is given by
\ber
{\overline {a}}_{non-res} & = & K \sqrt{m}\sqrt{m_1} m_1 \xi(v\cdot v_{+}) 
\frac{f_{D^*}}{f_K} \nonumber\\
& &\left[i\varepsilon^{\mu\nu\alpha\beta}\epsilon^{*}_{1\mu}
\epsilon^{*}_{2\nu}v_{\alpha}v_{+\beta} 
+\epsilon^{*}_{1}\cdot v \epsilon^{*}_{2}\cdot v_{+}-
\epsilon^{*}_{1} \cdot \epsilon^{*}_2(v\cdot v_{+} +1) \right] \
\eer
where
$K= \frac{G_F}{{\sqrt{2}}} V_c \left(\frac{c_1}{N_c} + c_2 \right) $.
Note that the amplitude above is the same as the amplitude for
$B^0 \to D^{*-} D^{*+} $ \cite{Rosner} except for a constant multiplicative 
factor 
$ \sim 1/f_K$. 

To a good approximation one can use $\vec{v} \sim 0$ where $\vec{v}$ is the velocity of the 
$ {\bar B}^0$ in the boosted frame where the two $D^*$ are moving back to back. 
The $K_s$, in this limit, is emitted in a s-wave configuration as the amplitude is
independent of the angles that specify the $K_s$ momentum in the boosted frame.
Then, as in the ${\bar{B}} ^0 \to D^{*+}D^{*-} $ case there are 
three helicity states 
 allowed, $(+,+)$, $(- ,-)$ and $(0,0)$ with the 
corresponding helicity amplitudes $H_{+ +}$, $H_{- -}$ and $H_{00}$. 
The helicity states are not CP eigenstates but one can go to 
the partial wave basis or the transverse basis 
where the states are  CP eigenstates.
The transverse basis amplitudes are related to the helicity amplitudes as
\ber
A_{\parallel} & = & \frac{H_{++} +H_{- -}}{\sqrt{2}}
\nonumber\\ 
A_{\perp} & = & \frac{H_{++} -H_{- -}}{\sqrt{2}}
\nonumber\\ 
A_0 & = & H_{0 0}\
\eer
The three partial waves that are allowed in this case, $s$, $p$ and $d$ are 
then given by
\ber
s=\frac{\sqrt{2}A_{\parallel} -A_0}{\sqrt{3}} \nonumber\\
p=A_{\perp} \nonumber\\
d=\frac{\sqrt{2}A_{0} +A_{\parallel}}{\sqrt{3}}
\eer
The CP of the final state is given by $\eta(-)^L$ where 
$\eta$ is the intrinsic 
parity of the final states 
and $L$ is the relative angular momentum between
$D^{*+}$ and $D^{*-}$.
In the approximation  $\vec{v} \sim 0$ one can write the non-resonant
 amplitude for
${ B}^0(v,m) \to D^{*+}(\epsilon_1,v_{+},m_1) 
D^{*-}(\epsilon_2,v_{-},m_1) K_s(p_k)$  
\ber
{{a}}_{non-res} & = & K \sqrt{m}\sqrt{m_1} m_1 \xi(v\cdot v_{-}) 
\frac{f_{D^*}}{f_K} \nonumber\\
& &\left[-i\varepsilon^{\mu\nu\alpha\beta}\epsilon^{*}_{2\mu}
\epsilon^{*}_{1\nu}v_{\alpha}v_{-\beta} 
+\epsilon^{*}_{2}\cdot v \epsilon^{*}_{1}\cdot v_{-}-
\epsilon^{*}_{1} \cdot \epsilon^{*}_2(v\cdot v_{-} +1) \right] \
\eer

There can also be pole contributions of the type shown in Figure 1. 

\begin{figure}[htb]
\centerline{\epsfysize 2.1 truein \epsfbox{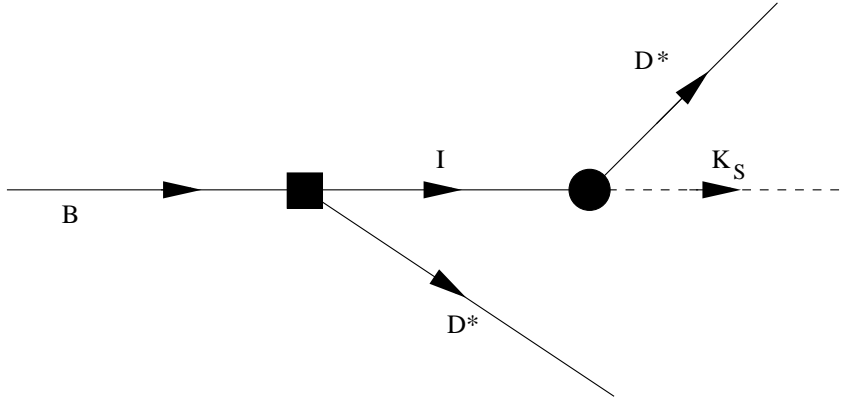}}
\caption{The pole contribution to the process
${ B}  \to D^{*}D^{*} K_S$. The intermediate state $I$ can be
 $ D^{ * \prime}_{s1}$ or  $D^{* }_{s}$. The solid square represents the 
weak vertex while the solid circle represents the strong vertex.}
%\label{bkx_qsquared}
\end{figure}

These give the decay sequences

$$ {\bar B}^0 \to D^{*+} D^{* \prime -}_{s1} \to D^{*+} D^{*-} K^0 $$ and 
$$ { \bar {B}}^0 \to D^{*+} D^{* -}_{s} \to D^{*+} D^{*-} K^0 $$
The propagator for the vector resonance is given by
\ber
S_{\mu \nu} & = & \frac{i(V_{\mu}V_{\nu} -g_{\mu \nu})}{2V\cdot k} \
\eer
where the momentum of the propagating particle $P =m_IV+k$ 
where $m_I$ is the mass of the intermediate particle in Figure. 1.

The contribution from the pole diagrams are given by ${\overline a}_{1res}$ and
${\overline a}_{2res}$, where ${ \overline a}_{1res}$ is,  
with $m_I=m^{* \prime}$,
\ber
{\overline {a}}_{1res} & = & K \sqrt{m}\sqrt{m_1} \sqrt{m_1} \sqrt{m^{*\prime}} \xi(v.v_{+}) 
\frac{f_{D^{* \prime}_{s1}}}{f_K}
\frac{h p_k\cdot v_{-}}{(p_k\cdot v_{-} +m_1-m^{*\prime} +
\frac{i\Gamma_{D^{*\prime}_{s1}}}{2})} \nonumber\\
& &\left[-i\varepsilon^{\mu\nu\alpha\beta}\epsilon^{*}_{1\mu}
\epsilon^{*}_{2\nu}v_{\alpha}v_{+\beta}
-\epsilon^{*}_{1}\cdot v \epsilon^{*}_{2}\cdot
v_{+}+\epsilon^{*}_{1}\cdot \epsilon^{*}_2(v\cdot v_{+} +1) . \right] \
\eer
Note that the above amplitude can be rewritten as
 \ber
{\overline {a}}_{1res} & = & -{\overline {a}}_{non-res} \frac{f_{D^{* \prime}_{s1}}}
{f_{D^*}}\sqrt{\frac{m^{* \prime}}{m_1}}
\frac{h p_k\cdot v_{-}}{(p_k\cdot v_{-} +m_1-m^{*\prime} +
\frac{i\Gamma_{D^{*\prime}_{s1}}}{2})} \
\eer
${\overline {a}}_{2res}$ is given by, with $m_I=m^*$ where $m^*$ is the $1^{-}$ $D_s^*$ mass,
\ber
{\overline {a}}_{2res} & =&  K \sqrt{m}\sqrt{m_1} 
\sqrt{m_1} \sqrt{m^*} \xi(v.v_{+}) 
\frac{f_{D^{*}_s}}{f_K}
\frac{g }{(p_k\cdot v_{-} +(m_1-m^*) +\frac{i\Gamma_{D_s^*}}{2})}X \nonumber\\
X &= &-i\varepsilon^{\mu\nu\alpha\beta}\epsilon^{*}_{2\mu}
p_{k\nu}v_{+\alpha}v_{-\beta}\epsilon^{*}_1.v
+i\varepsilon^{\mu\nu\alpha\beta}\epsilon^{*}_{1\mu}
\epsilon^{*}_{2\nu}p_{k \alpha}v_{-\beta}(v\cdot v_{+} +1)
+(\epsilon^{*}_{1}\cdot v_{-} \epsilon^{*}_{2}\cdot v p_k\cdot v_{+}-
\epsilon^{*}_{1}\cdot v_{-} \epsilon^{*}_{2}\cdot v_{+} p_k\cdot v) \nonumber\\
&+&(\epsilon^{*}_{1}\cdot p_k \epsilon^{*}_{2}\cdot v_{+} v\cdot v_{-}-
\epsilon^{*}_{1}\cdot p_k \epsilon^{*}_{2}\cdot v v_{+}\cdot v_{-})
+(\epsilon^{*}_{1}\cdot \epsilon^{*}_{2} p_k\cdot v v_{+}\cdot v_{-}-
\epsilon^{*}_{1}\cdot \epsilon^{*}_{2} p_k\cdot v_{+} v\cdot v_{-}) \
\eer

The amplitude ${ \overline {a}}_{2res}$ gives a tiny contribution to the total amplitude and can 
be neglected. In fact, this amplitude vanishes in the small velocity limit 
where the $D^*$ are almost at rest \cite{Shifman}. We note that the process
with the $0^+$ intermediate state
$$ { \bar {B}}^0 \to D^{*+} D^{- }_{s0} \to D^{*+} D^{*-} K^0 $$ is 
not allowed 
due to parity conservation while the amplitude with the $0^-$ 
intermediate state
$$ { \bar {B}}^0 \to D^{*+} D^{- }_{s} \to D^{*+} D^{*-} K^0 $$ is expected to be small 
compared to ${ \overline {a}}_{1res}$.
The propagator term in the above amplitude goes as approximately 
$1/(E_K + (m_{D^*}-m_{D_s}))$ which does not have a pole as in
${\overline {a}}_{1res}$. Moreover,
the amplitude is further suppressed with respect to 
${\overline{a}}_{1res}$ by a 
factor $ \sim p_k/E_k$ or $|\vec{v}|/v_0$, where $\vec{v}$ and $v_0$ are 
the three velocity and the time component of the velocity four vector of 
the $D^*$,  from  the
$ D^{+ }_{s}  D^{*+} K^0 $ vertex.

The total amplitude for ${\bar B}^0(v,m) \to D^{*+}(\epsilon_1,v_{+},m_1) 
D^{*-}(\epsilon_2,v_{-},m_1) K_s(p_k)$ can be written as
\ber
{\overline a} & = & {\overline {a}}_{non-res}[1-P_1]\
\eer 
and the total amplitude  for
${ B}^0(v,m) \to D^{*+}(\epsilon_1,v_{+},m_1) D^{*-}(\epsilon_2,v_{-},m_1) 
K_s(p_k)$   
can be written as
\ber
{a} & = & { {a}}_{non-res}[1-P_2]\
\eer 
with
\ber
P_1 & = & \frac{f_{D^{* \prime}_{s1}}}{f_{D^*}}\sqrt{\frac{m^{*\prime}}{m_1}}
\frac{h p_k\cdot v_{-}}{(p_k\cdot v_{-} +m_1-m^{*\prime} +
\frac{i\Gamma_{D^{*\prime}_{s1}}}{2})} \\
P_2 & = & \frac{f_{D^{* \prime}_{s1}}}{f_{D^*}}\sqrt{\frac{m^{*\prime}}{m_1}}
\frac{h p_k\cdot v_{+}}{(p_k\cdot v_{+} +m_1-m^{*\prime} +
\frac{i\Gamma_{D^{*\prime}_{s1}}}{2})} \
\eer
  
Note $P_1$ and $P_2$ can be expressed in terms of $E_k$ 
and $y$ and  $P_1(y, E_k) =P_2(-y,E_k)$.
The relation between quantities in the boosted frame and the rest frame of 
the $B$ along with the calculation of the squared amplitude are given in Appendix B.

The double differential decay distribution for the time independent
process
$${\bar B}^0(v,m) \to D^{*+}(\epsilon_1,v_{+},m_1) D^{*-}(\epsilon_2,v_{-},m_1) K_s(p_k)$$ 
can be written as
\ber
\frac{1}{\Gamma}\frac{d\Gamma}{dy dE_k} & =& \frac{f(y,E_k)}
{\int f(y,E_k) \frac{p_{k}^{\prime}p_{+}^{\prime}}{m}dydE_k}\
\eer
where $p_{k}^{\prime}$ and $p_{+}^{\prime}$ are 
the magnitudes of the three momentum of the 
$K_s$ and $D^{*+}$ in the boosted frame and the expression for $f(y,E_k)$ can be found in Appendix B.
The differential distribution depends only on
$\frac{f_{D^{* \prime}_{s1}}}{f_{D^{*}}}$,
the mass $m^{* \prime}$ and
 the coupling $h$ of the $D_{s1}^{* \prime}$ state . It is expected that
${f_{D^{* \prime}_{s1}}} \approx f_{D^{*}_s}$ and in the $SU(3)$ limit
$f_{D^{*}_s}=f_{D^{*}}$.
 So in the $SU(3)$ limit
 a two parameter fit to the 
differential decay distribution can determine the mass
and the coupling of the
$D_{s1}^{* \prime}$ state .
 
The widths of the positive parity excited states are expected to be 
saturated by single kaon transitions \cite{Fluke}. In our calculation we 
require the width of the $D_{s1}^{* \prime}$ state.
Assuming
\ber
\Gamma_{D^{*+\prime}_{s1}} & \approx & 
\Gamma( D^{*+\prime}_{s1} \to D^{*+} K^0)
+ \Gamma( D^{*+\prime}_{s1} \to D^{*0} K^+)\
\eer
one can write
\ber
\Gamma_{D^{*\prime}_{s1}} & = & \frac{h^2}{\pi f_K^2}
\frac{m_1}{m^{* \prime}}
(m^{*\prime} -m_1)^2 p \
\eer
where $p$ is the magnitude of the three momentum of the decay 
products in the rest frame
of $ D_{s1}^{* \prime}$ and $m_1$ and $m^{*\prime}$ 
are the masses of the $D_s^*$ and
 $D_{s1}^{* \prime}$ state .

It is clear that if $a= {\bar a}$ then the dilution factor $D=1$. However 
that is not 
the case here. For the non resonant contribution, 
in the approximation of small 
velocity of the $B$, the final state is an admixture of CP states 
with different CP parities.
This leads to $D<1$. This is the same dilution of the asymmetry 
as in the case for
$B \to D^{*+} D^{*-}$. When the resonant contribution is included 
 the amplitude $a$ and ${\bar a}$ have an 
asymmetric dependence on the 
variable $y$. This reflects the fact that in the  process
$ {\bar B}^0 \to D^{*+} D^{* \prime -}_{s1} \to D^{*+} D^{*-} K^0 $ 
the kaon emerges 
most of the time closer to $D^{*-}$ than the $D^{*+}$. The situation 
is reversed
for $B^0$ decays. Consequently there is additional mismatch between the 
amplitudes $a$ and ${\bar a}$  which leads to further dilution of the 
asymmetry.
One can reduce the 
dilution of the asymmetry, { \it i.e.}, increase $D$,
by imposing cuts so as to reduce the resonant contribution. 
We consider several  cases where cuts may be employed to decrease 
the dilution of the asymmetry.
 \null From Eq. (31-32) it is clear that resonance occurs when the 
following condition is met
\ber
p_k \cdot v_{+}  &= & m^{*'}-m_1 \\
p_k \cdot v_{-}  &= & m^{*'}-m_1 \
\eer

If, in the allowed region of $E_k$, we can find a value $E_{k0}$ such that
for values of $E_k \ge E_{k0}$ the above conditions are not satisfied for 
$ -1 \le y \le 1$ 
then we can remove the resonance by using  the cut $E_k \ge E_{k0}$ . The 
value of $E_{k0} \sim 0.76$ GeV in our case. We will call this case cut 
1 for future reference. 

Another possible cut is to include the whole range of $E_k$ but in the region
$E_k \le E_{k0}$ we remove the resonance by cutting on the variable $y$. 
We can use the region
$ -0.5 \le y \le 0.5$ since for most values of $E_k$ the 
resonance condition is satisfied in the range
$ -1 \le y \le -0.5$  and $ 0.5 \le y \le 1$.  
 We will call this case cut 
2 for future reference. In any event, the cuts can be optimized after
the resonance has been seen experimentally.
However as we 
try to increase the value of $D$ by reducing the resonance through cuts we 
also lessen the branching ratio from loss of signal.  

\section{Results and Discussions}
\begin{figure}[htb]
\centerline{\epsfysize 3.2 truein \epsfbox{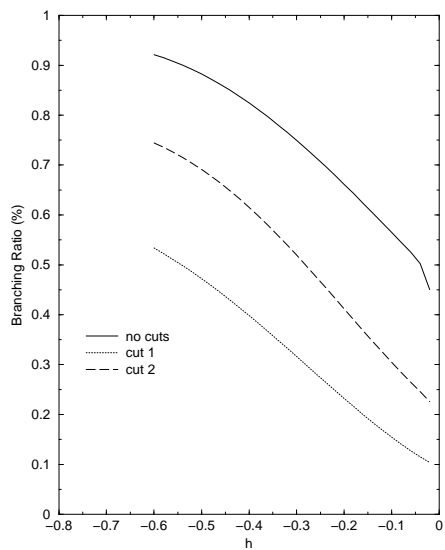}}
\caption{The branching fraction for 
${\bar B}^0 \to D^{*+} 
D^{*-} K_s$  as a function of the  $h$ with and without cuts. }
%\label{amp-ampbar}
\end{figure}

\begin{figure}[htb]
\centerline{\epsfysize 3.2 truein \epsfbox{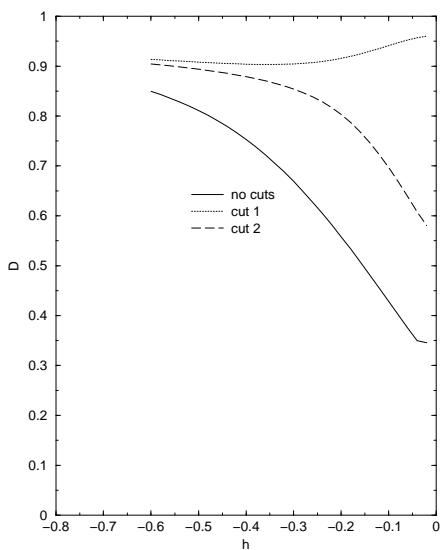}}
\caption{The Dilution factor $D$
  as a function of the  $h$ with and without cuts. } 
%\label{amp-ampbar}
\end{figure}

\begin{figure}[htb]
\centerline{\epsfysize 3.2 truein \epsfbox{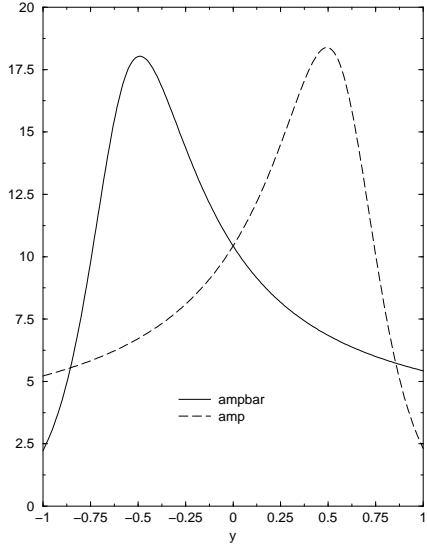}}
\caption{The squared amplitude for ${ B}^0 \to D^{*+} 
D^{*-} K_s$   
${\bar B}^0 \to D^{*+} 
D^{*-} K_s$  as a function of the variable $y$ for $h=-0.4$ which corresponds
to a $D_{s1}^{* \prime}$ state  with a width of about 150 MeV. }
%\label{amp-ampbar}
\end{figure}

\begin{figure}[htb]
\centerline{\epsfysize 3.2 truein \epsfbox{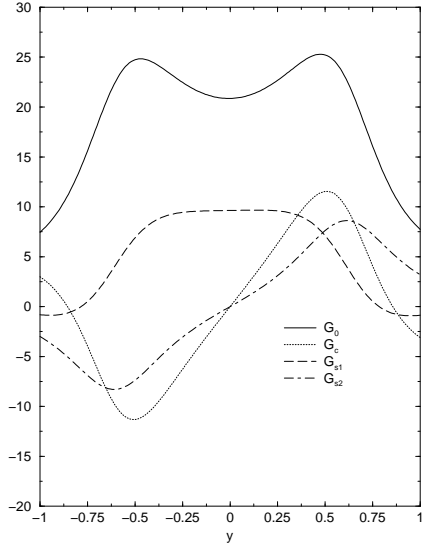}}
\caption{The function 
$G_0$,$G_c$, $G_{s1}$ and $G_{s2}$ as a function of $y$ for $E_k=0.6 $ GeV and
$h=-0.4$ }
\end{figure}

\begin{figure}[htb]
\centerline{\epsfysize 3.2 truein \epsfbox{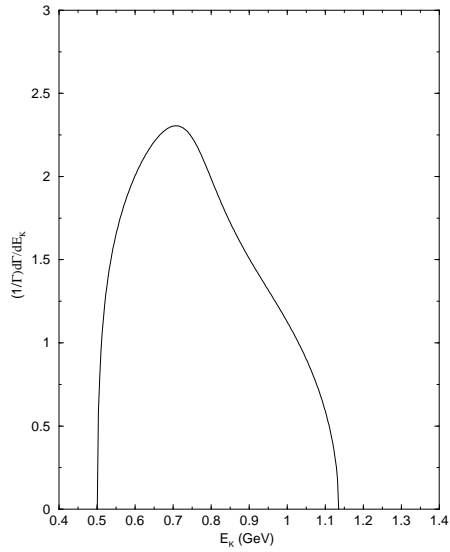}}
\caption{The decay distribution $d\Gamma/dE_k$ versus the kaon energy
$E_k$.  }
%\label{decaydist}
\end{figure}

\begin{figure}[htb]
\centerline{\epsfysize 3.2 truein \epsfbox{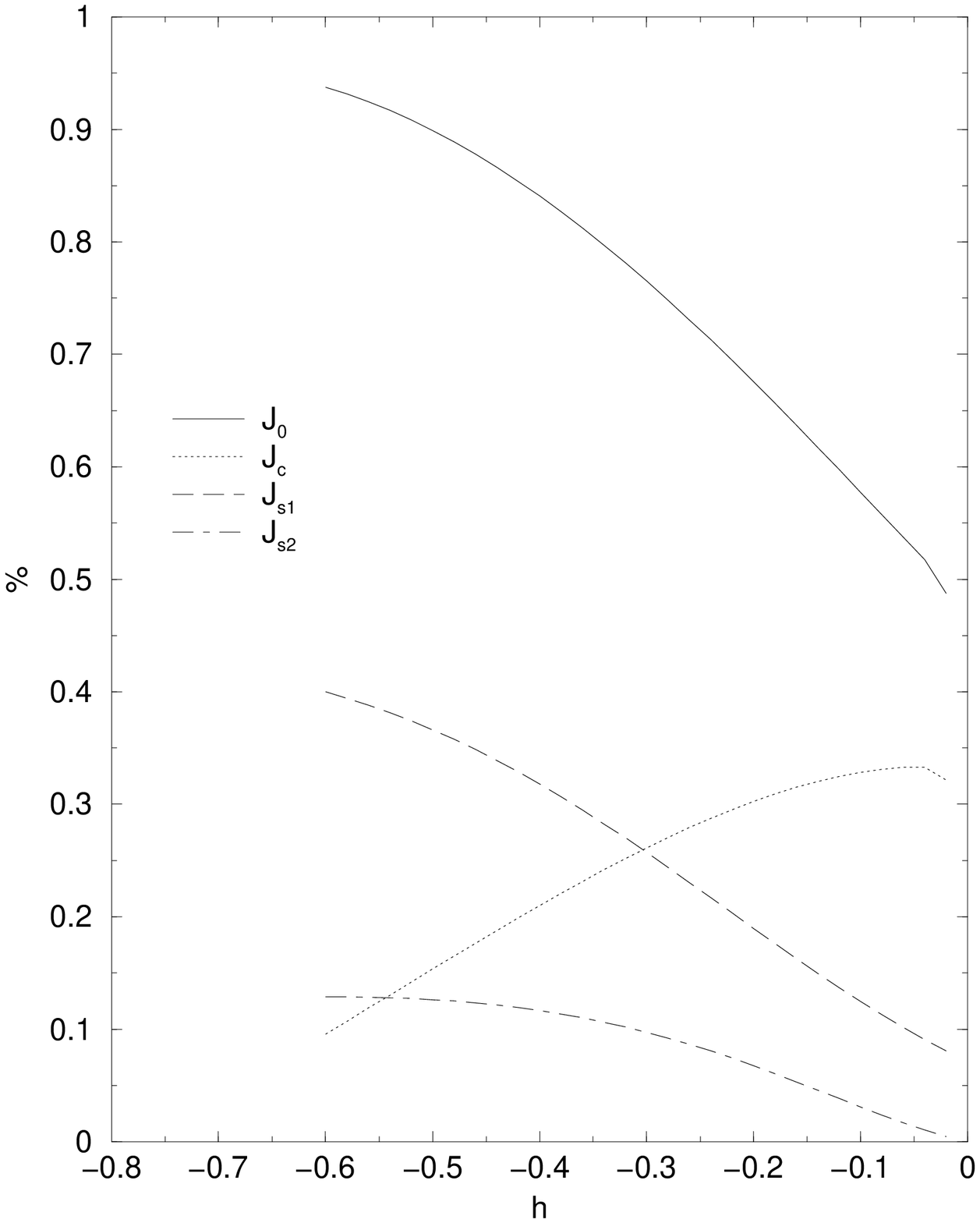}}
\caption{The functions 
$G_0$,$G_c$, $G_{s1}$ and $G_{s2}$ integrated over the
 $y \ge 0$ as a function of 
$h$. ${J}_0$ and ${J}_c$ refer to the integrated
$G_0$ and $G_c$ functions while $ J_{s1}$ and $ J_{s2}$ 
refer to the integrated
 $G_{s1}$ and $G_{s2}$ functions. The values of the integral can be 
obtained by multiplying by $ \Gamma_{B}$ where 
$\Gamma_{B}$ is the width of the $B$. }
\end{figure}

 As inputs to the calculation, we use 
$f_{D_s^*} \approx f_{D^{* \prime}_{s1}} =200$ MeV  and take the mass of the
 $D_{s1}^{* \prime}$ state to be $ 2.6$ GeV. For the Isgur-Wise function
 we use the form
$$ \xi (\omega) = (\frac{2}{1+ \omega})^2. $$ 
 QCD sum rules have been used to compute the strong
coupling constants $g$  and $h$ \cite{QCDsumrule}. 
We will use $g=0.3$ as obtained in Ref \cite{QCDsumrule} 
but keep $h$ as a free parameter because this coupling 
plays a more important role in the decay widths.

Fig. 2 shows the branching fraction for ${ \bar B}^0 \to D^{*+} D^{*-} K_s$ 
 as a function of the coupling $h$. A QCD sum rule calculation gives
$h \sim -0.5$ \cite{QCDsumrule}. We use the same sign of 
$h$ as obtained in QCD sum rule calculation but 
vary $h$ from $-0.6 $ to $-0.1$. 
For this range of $h$ the branching fraction
can vary in the range $0.45-0.93 \%$ when we employ no cuts. 
For $h=-0.4$ which corresponds
to a $D_{s1}^{* \prime}$ state  
with a width of about 150 MeV
the branching fraction is $ 0.83 \%$.  
In our calculation 
this corresponds to a 
branching ratio
${\cal B}({{B}}^0\to D^{*-} {D}^{*+} K^0) \approx 
{\cal B}({ {B}}^0\to D^{*-} {D}^{*0} K^+) \approx 
{\cal B}(B^+ \to \bar{D}^{*0} {D}^{*0} K^+) \approx
 {\cal B}(B^+ \to \bar{D}^{*0} D^{*+} K^0) \approx 0.9-1.86 \%$. This 
is consistent with the CLEO measurements mentioned above. Fig. 2 also shows  
 the branching fractions after cuts have been applied
 to reduce 
the dilution of the CP asymmetry.

Fig. 3 shows a plot of the dilution factor $D$ versus the coupling $h$. In 
the absence of any cuts we find that
larger values of $|h|$ gives a larger value of $D$ and 
hence less dilution in the asymmetry because for a broad
$D_{s1}^{* \prime}$ state   there is more overlap between the amplitudes for
${ B}^0 \to D^{*+} 
D^{*-} K_s$ and   
${\bar B}^0 \to D^{*+} 
D^{*-} K_s$.  
 For $ h=-0.4$ the dilution factor is about $0.75$ 
with no cuts. For the case of cut 1, where we use the cut $E_k>E_{k0}$
 to effectively remove the resonance, the dilution 
factor increases  with smaller $|h|$. This is because for 
smaller $|h|$ and $E_k>E_{k0}$ the resonant amplitude is small 
and the total amplitude is dominated by 
the non-resonant amplitude which give a larger value for $D$. 
For the case of cut 2, as in the case with no cuts, the dilution factor 
$D$ decreases with smaller $|h|$. This is because we are using the 
entire region of $E_k$ and not removing the resonance by the cut
$E_k>E_{k0}$ as in the case of cut 1. Consequently a broader resonance and hence a larger value of $|h|$ gives a larger value of $D$ and vice versa.  

Figure 4 shows the squared amplitude for ${ B}^0 \to D^{*+} 
D^{*-} K_s$ and   
${\bar B}^0 \to D^{*+} 
D^{*-} K_s$  as a function of the variable $y$ for $h=-0.4$. As mentioned 
above the  nature of the two curves
 reflects the fact that in the  process
$ {\bar B}^0 \to D^{*+} D^{* \prime -}_{s1} \to D^{*+} D^{*-} K^0 $ 
the kaon emerges 
most of the time closer to $D^{*-}$ than the $D^{*+}$ while the situation 
is reversed
for $B^0$ decays.

Fig. 5 shows the plot of the function 
$G_0$,$G_c$, $G_{s1}$ and $G_{s2}$ as a function of $y$
for $E_k=0.6$ GeV and for $h=-0.4$. From the figure 
we see that the functions  
$G_0$ and $G_{s1}$ are symmetric in $y$ while
$G_c$ and $G_{s2}$ are antisymmetric in $y$. This follows from 
the absence of direct CP violation as shown in Eq. (19-22).

In Fig. 6 we show the 
the decay distribution $d\Gamma/dE_k$ versus the kaon energy
$E_k$. 
For small values of $E_k$ the decay distribution 
shows a clear resonant structure which comes from the pole 
contribution to $a_{1res}$ with the excited $J^P=1^+$
 intermediate state. Therefore, 
examination of the $D^{*} K_s$ mass spectrum
may be the best experimental way to find the broad $1^+$ p-wave
$D_s$ meson and as mentioned in the previous section a fit to the
decay distribution will measure its mass and the coupling.

 In Fig. 7 we show the functions 
$G_0$, $G_c$, $G_{s1}$ and $G_{s2}$ integrated over the
 $y \ge 0$ as a function of 
$h$. ${J}_0$ and ${J}_c$ refer to the integrated
$G_0$ and $G_c$ functions while $ J_{s1}$ and $ J_{s2}$ 
refer to the integrated
 $G_{s1}$ and $G_{s2}$ functions. As already mentioned, restricting 
the integration range to 
$y \ge 0$
allows a probe of the $\cos( 2 \beta)$ term in the time dependent rate
for $B^0 (t) \to D^{(*+)}{ D}^{(*-)}K_s$ decays. It is clear from 
the figure that a broader resonance is more favorable to probe
$G_{s2}$ which is the coefficient of the $\cos (2 \beta)$ term.   

In summary, we have studied the possibility of extracting $\sin (2 \beta)$ 
 and $ \cos ( 2 \beta)$ from time dependent
 $B^0 \to D^{(*)}{\overline D}^{(*)}K_s$ decays.  These decays are 
expected to
have less penguin contamination and much larger
 branching fractions than the two body
modes  $B^0 \to D^{(*)}{\overline D}^{(*)}$ 
. Using HHCHPT we have calculated the branching fractions and the various 
coefficient functions that appear in the time dependent rate for
$B^0 \to D^{(*+)}{ D}^{(*-)} K_s$. We also showed that a
examination of the $D^{*} K_s$ mass spectrum
may be the best experimental way to find the broad $1^+$ p-wave
$D_s$ meson and measure its mass and coupling.

\section{Acknowledgements}

We thank Yuval Grossman for an important observation. This work was
supported in part by the United States Department of Energy 
(T. Browder and S. Pakvasa),
 and by the Natural  Sciences and Engineering Research  Council
of Canada (A. Datta and P. J. O'Donnell).  
\newpage
\setcounter{equation}{0}
\def\theequation {A. \arabic {equation}}
\section*{Appendix A}
In the Standard Model (SM) 
the amplitudes for $ B \to D^{(*)} {\bar D}^{(*)} K_s$
are generated by the following effective 
Hamiltonian \cite{Reina,buras}:
\begin{eqnarray}
H_{eff}^q &=& {G_F \over \protect \sqrt{2}} 
   [V_{fb}V^*_{fq}(c_1O_{1f}^q + c_2 O_{2f}^q) -
     \sum_{i=3}^{10}(V_{ub}V^*_{uq} c_i^u
+V_{cb}V^*_{cq} c_i^c +V_{tb}V^*_{tq} c_i^t) O_i^q] +H.C.\;,
\end{eqnarray}
where the
superscript $u,\;c,\;t$ indicates the internal quark, $f$ can be $u$ or 
$c$ quark, $q$ can be either a $d$ or a $s$ quark depending on 
whether the decay is a $\Delta S = 0$
or $\Delta S = -1$ process.
The operators $O_i^q$ are defined as
\begin{eqnarray}
O_{1f}^q &=& \bar q_\alpha \gamma_\mu Lf_\beta\bar
f_\beta\gamma^\mu Lb_\alpha\;,\;\;\;\;\;\;O_{2f}^q =\bar q
\gamma_\mu L f\bar
f\gamma^\mu L b\;,\nonumber\\
O_{3,5}^q &=&\bar q \gamma_\mu L b
\bar q' \gamma_\mu L(R) q'\;,\;\;\;\;\;\;\;O_{4,6}^q = \bar q_\alpha
\gamma_\mu Lb_\beta
\bar q'_\beta \gamma_\mu L(R) q'_\alpha\;,\\
O_{7,9}^q &=& {3\over 2}\bar q \gamma_\mu L b  e_{q'}\bar q'
\gamma^\mu R(L)q'\;,\;O_{8,10}^q = {3\over 2}\bar q_\alpha
\gamma_\mu L b_\beta
e_{q'}\bar q'_\beta \gamma_\mu R(L) q'_\alpha\;,\nonumber
\end{eqnarray}
where $R(L) = 1 \pm \gamma_5$, 
and $q'$ is summed over all flavors except t.  $O_{1f,2f}$ are the 
current-current operators that represent tree level processes. $O_{3-6}$ are the strong gluon induced
penguin operators, and operators 
$O_{7-10}$ are due to $\gamma$ and Z exchange (electroweak penguins),
and ``box'' diagrams at loop level. The Wilson coefficients
 $c_i^f$ are defined at the scale $\mu \approx m_b$ 
and have been evaluated to next-to-leading order in QCD.
The $c^t_i$ are the regularization scheme 
independent values obtained in Ref. \cite{FSHe}.
We give the non-zero  $c_i^f$ 
below for $m_t = 176$ GeV, $\alpha_s(m_Z) = 0.117$,
and $\mu = m_b = 5$ GeV,
\begin{eqnarray}
c_1 &=& -0.307\;,\;\; c_2 = 1.147\;,\;\;
c^t_3 =0.017\;,\;\; c^t_4 =-0.037\;,\;\;
c^t_5 =0.010\;,
 c^t_6 =-0.045\;,\nonumber\\
c^t_7 &=&-1.24\times 10^{-5}\;,\;\; c_8^t = 3.77\times 10^{-4}\;,\;\;
c_9^t =-0.010\;,\;\; c_{10}^t =2.06\times 10^{-3}\;, \nonumber\\
c_{3,5}^{u,c} &=& -c_{4,6}^{u,c}/N_c = P^{u,c}_s/N_c\;,\;\;
c_{7,9}^{u,c} = P^{u,c}_e\;,\;\; c_{8,10}^{u,c} = 0
\end{eqnarray}
where $N_c$ is the number of color. 
The leading contributions to $P^i_{s,e}$ are given by:
 $P^i_s = ({\frac{\alpha_s}{8\pi}}) c_2 ({\frac{10}{9}} +G(m_i,\mu,q^2))$ and
$P^i_e = ({\frac{\alpha_{em}}{9\pi}})
(N_c c_1+ c_2) ({\frac{10}{9}} + G(m_i,\mu,q^2))$.  
The function
$G(m,\mu,q^2)$ is given by
\begin{eqnarray}
G(m,\mu,q^2) = 4\int^1_0 x(1-x)  \mbox{ln}{m^2-x(1-x)q^2\over
\mu^2} ~\mbox{d}x \;.
\end{eqnarray}
All the above coefficients are obtained up to one loop order in electroweak 
interactions. The momentum $q$ is the momentum carried by the virtual gluon in
the penguin diagram.
When $q^2 > 4m^2$, $G(m,\mu,q^2)$ becomes imaginary. 
In our calculation, we 
use $m_u = 5$ MeV, $m_d = 7$ MeV, $m_s = 200$ MeV, $m_c = 1.35$ GeV
\cite{lg,PDG}.

In the factorization assumption the amplitude for 
$ B \to D^{(*)} {\overline D}^{(*)} K_s$ can now be written as
\be
	M = M_1 + M_2 +M_3 +M_4
\ee
where
\ber
	M_1 & = &\frac{G_F}{\protect \sqrt{2}} X_1
 <{\overline {D}}^{(*)}K_s|\, \bar{s} \gamma^\mu(1-\gamma^5) \, c\, |\, 0>
	       <D^{(*)} |\, \bar{c} \, \gamma_\mu (1-\gamma^5) \, b\, |\, B>
\nonumber  \\
M_2 & = &\frac{G_F}{\protect \sqrt{2}} X_2
 <{\overline {D}}^{(*)} D^{(*)}|\, \bar{c} \gamma^\mu(1-\gamma^5) \, c\, |\, 0>
	       <K_s |\, \bar{s} \, \gamma_\mu (1-\gamma^5) \, b\, |\, B>
\nonumber  \\
M_3 & = &\frac{G_F}{\protect \sqrt{2}} X_3
 <{\overline {D}}^{(*)} D^{(*)}|\, \bar{c} \gamma^\mu(1+\gamma^5) \, c\, |\, 0>
	       <K_s |\, \bar{s} \, \gamma_\mu (1-\gamma^5) \, b\, |\, B>
\nonumber  \\
M_4 & = &\frac{G_F}{\protect \sqrt{2}} X_4
 <{\overline {D}}^{(*)}K_s|\, \bar{s} (1+\gamma^5) \, c\, |\, 0>
	       <D^{(*)} |\, \bar{c} \, (1-\gamma^5) \, b\, |\, B>  \\
\eer
where
\ber
        X_1 & = & V_c \left(\frac{c_1}{N_c} + c_2 \right) + 
                  \frac{B_3}{N_c} + B_4
		+ \frac{B_9}{N_c} + B_{10} \nonumber\\
	X_2 & = & V_c \left(c_1 + \frac{c_2}{N_c} \right) + B_3 
		+ \frac{1}{N_c} B_4 + B_9 + \frac{1}{N_c} B_{10}\nonumber  \\
        X_3 & = & B_5 + \frac{1}{N_c} B_6 + B_7 + \frac{1}{N_c} B_8 \nonumber \\
        X_4 & = & -2 \, \left(\frac{1}{N_c} B_5 + B_6 + \frac{1}{N_c} B_7 
		+ B_8 \right)
\eer
We have defined
\be
	B_i = - \sum_{q=u,c,t} c_i^q V_q
\ee
with
\be
	V_q = V_{qs}^{*} V_{qb}
\ee

In the above equations $N_c$ represents the number of colors. 
It is usually the practice in the study of two body non-leptonic 
decays to include non-factorizable effects by the 
replacement $N_c \to N_{eff}$. 
Since it is not obvious that $N_{eff}$ for 
two body non-leptonic decays is the same for 
non-leptonic three body decays we will use $N_c=3$ in our calculation.
 
As already mentioned, we expect the contribution from penguin
diagrams to be small
and so as a first approximation we will neglect $M_3$ and
$M_4$. Furthermore, from the 
values of the Wilson coefficients $c_{1,2}$ given above in the previous
section it is clear that the amplitude $M_2$ is
suppressed with respect to $M_1$ with the Wilson coefficients associated with 
 $M_2$ being about 7 $\%$ of the
Wilson coefficients associated with  $M_1$.
We also note that the currents 
$<{\overline {D}}^{(*)}K_s|\, \bar{s} \gamma^\mu(1-\gamma^5) \, c\, |\, 0>$ and
$<K_S |\, \bar{s} \, \gamma_\mu (1-\gamma^5) \, b\, |\, B>$, 
which appear in $M_1$ and $M_2$ respectively,
receive contributions from both the contact terms 
and the pole terms. For the former current 
the pole terms are proportional to
$1/(E_K - \delta m)$ 
while for the latter the pole term goes as $1/(E_K + \delta m)$. 
This also leads to a further
suppression of $M_2$ relative to $M_1$. We therefore neglect 
$M_2$ and only retain $M_1$ in our calculation. We will also neglect CP 
violation in the $K^0-\bar{K}^0$ system and so
 (with an appropriate choice of phase
convention) we can write 
$$K_s = \frac{ K^0 - {\overline K}^0 }{\sqrt{2}} $$

To calculate the various matrix elements in $M_1$ above we use
 Heavy Hadron Chiral Perturbation Theory (HHCHPT). In HHCHPT, the ground state
$(j^P \, = \; {\frac{1}{2}}^-)$ heavy mesons  are described by the
 $4 \times 4$ Dirac matrix
\begin{equation}
H_a = \frac{(1+\rlap/v)}{2}[P_{a\mu}^*\gamma^\mu-P_a\gamma_5]
\label{h}
\end{equation}
where $v$ is the heavy meson velocity,
$P^{*\mu}_a$ and $P_a$ are annihilation operators
of the $1^-$ and $0^-$ $Q{\bar q}_a$ mesons
($a=1,2,3$ for $u,d$ and $s$): for charm, they are $D^*$ and $D$
respectively.
 The  field ${\overline H}_a$ is defined by
$$ {\overline H}_a =\gamma^0 H^{\dagger}\gamma^0 $$
Similarly, the positive parity $1^+$ and $0^+$ states
$(j^P \, = \; {\frac{1}{2}}^+)$
are described by
\begin{equation}
S_a=\frac{(1+\rlap/v)}{2} \left[D^{*\prime}_{1 \mu}\gamma^\mu\gamma_5-D_0\right] \;.
\label{s}
\end{equation}

In the above equations 
$v$ generically represents the heavy meson four-velocity and 
 $D^{*\mu}$ and $D$ 
are annihilation
operators normalized as follows:
\begin{eqnarray}
\langle 0|D| c{\bar q} (0^-)\rangle & =&\sqrt{M_H} \nonumber\\
\langle 0|{D^*}^\mu| c{\bar q} (1^-)\rangle & = & \epsilon^{\mu}\sqrt{M_H}.
\end{eqnarray}
 Similar equations hold for the positive parity states 
$D_{1\mu}^{*\prime}$ and $D_0$ .
The vector states in the multiplet satisfy
the transversality conditions 
$$v^\mu D^*_{\mu}=v^\mu D_{1\mu}^{*\prime}= 0.$$

For the octet of the pseudo Goldstone bosons, one uses the exponential form:
\begin{equation}
\xi=\exp{\frac{iM}{f_{\pi}}}
\end{equation}
where
\begin{equation}
{M}=
\left (\begin{array}{ccc}
\sqrt{\frac{1}{2}}\pi^0+\sqrt{\frac{1}{6}}\eta & \pi^+ & K^+\nonumber\\
\pi^- & -\sqrt{\frac{1}{2}}\pi^0+\sqrt{\frac{1}{6}}\eta & K^0\\
K^- & {\bar K}^0 &-\sqrt{\frac{2}{3}}\eta
\end{array}\right )
\end{equation}
and $f_{\pi}=132 \; MeV$.

The lagrangian describing the fields $H$, $S$ and $\xi$ and their interactions,
under the hypothesis of chiral and spin-flavor symmetry and at the lowest
order in light mesons derivatives, is\cite{Fluke}:
\begin{eqnarray}
{\cal L} &=& \frac{f_{\pi}^2}{8}
Tr\left[\partial^\mu\Sigma\partial_\mu
\Sigma^\dagger  +i  H_b v^\mu D_{\mu ba} {\bar H}_a \right]  \nonumber\\
& + & Tr \left[ S_b \;( i \; v^\mu D_{\mu ba} \; - \; \delta_{ba} \; \Delta)
{\bar S}_a \right] +\; i \, g Tr \left[H_b \gamma_\mu \gamma_5 
{\cal A}^\mu_{ba}
{\bar H}_a\right] \nonumber\\
& + & i \, g' Tr \left[S_b \gamma_\mu \gamma_5 {\cal A}^\mu_{ba} {\bar S}_a\right]
+ \, i \, h Tr\left[S_b \gamma_\mu \gamma_5 {\cal A}^\mu_{ba} {\bar H}_a 
\right] \
+ \; h.c. \label{L}
\end{eqnarray}
where $Tr$ means the trace, and
\begin{eqnarray}
D_{\mu ba}&=&\delta_{ba}\partial_\mu+{\cal V}_{\mu ba}
=\delta_{ba}\partial_\mu+\frac{1}{2}\left(\xi^\dagger\partial_\mu \xi
+\xi\partial_\mu \xi^\dagger\right)_{ba}\\
{\cal A}_{\mu ba}&=&\frac{1}{2}\left(\xi^\dagger\partial_\mu \xi-\xi
\partial_\mu \xi^\dagger\right)_{ba} \\
\end{eqnarray}
$\Sigma= \xi^2$ and $\Delta$ is the mass splitting of the $S_a$ states from
the ground state $H_a$.

The currents involving the heavy $b$ and $c$ quarks,
 $J^{\mu}_{V} =<D^{*}(\epsilon_1,p_1)|{\overline c}
\gamma^{\mu}(1-\gamma_5)b|B(p)> $
can be expressed in 
general in terms of form factors \cite{BSW}
\ber
J^{\mu}_{V} & = & \frac{-2iV(q^2)}{m+m_1}\varepsilon^{\mu\nu\alpha\beta}
\epsilon^{*}_{1\nu}p_{\alpha}
p_{1\beta}
-(m+m_1)A_1(q^2)\epsilon^{* \mu}_{1} 
+\frac{A_2(q^2)}{m+m_1}\epsilon^{*}_1 \cdot q(p +p_1)^{\mu} \nonumber\\
&+&2m_1A_3(q^2)\frac{\epsilon^{*}_1\cdot q}{q^2}q^{\mu}
-2m_1A_0(q^2)\frac{\epsilon^{*}_1 \cdot q}{q^2}q^{\mu} \
\eer
with
\ber
A_3(q^2) & = & \frac{m+m_1}{2m_1}A_1(q^2) - \frac{m-m_1}{2m_1}A_2(q^2) 
\nonumber\\
A_3(0) & =& A_0(0) \
\eer
where $q=p-p_1$ is the momentum transfer and $m$ and $m_1$ are the masses of $B$ and $D^{*}$.
In the heavy quark limit the various form factors are related to a universal Isgur-Wise function
$\xi(v\cdot v_1)$ where $v$ and $v_1$ are the four velocities of the $B$ and the $D^{*}$ meson.
One can write
\ber
J^{\mu}_{V} & = & \sqrt{m}\sqrt{m_1}\xi(v\cdot v_1)
\left[-i\varepsilon^{\mu\nu\alpha\beta}
\epsilon^{*}_{1\nu}v_{\alpha}v_{1\beta}
+v^{\mu}_{1}\epsilon^{*}_1\cdot v -\epsilon^{*\mu}_{1}(v\cdot v_1 +1) \right] \
\eer

 The  
 weak current 
$L^{\mu}_a={\overline q}^a \gamma^{\mu}(1-\gamma_5) Q$ can be written 
in the effective theory as
\ber
L^{\mu}_a & = & \frac{if_H\sqrt{m_H}}{2} Tr\left[\gamma^{\mu}(1-\gamma_5)H_b \xi^{+}_{ba}
\right] \
\eer
where $f_Q $ is the heavy meson decay constant.
 One can therefore write
\ber
<{\overline D}^{*}(\epsilon_2,v_2) {\overline K}^0
| {\overline s} \gamma_{\mu}(1-\gamma_5)c|0>
& = & i\frac{ {m_2} f_{D^*}\epsilon^*_{2 \mu}}{f_K} \
\eer

\setcounter{equation}{0}
\def\theequation {B. \arabic {equation}}
\section*{Appendix B}
The total amplitude for ${\bar B}^0(v,m) \to D^{*+}(\epsilon_1,v_{+},m_1) 
D^{*-}(\epsilon_2,v_{-},m_1) K_s(p_k)$ can be written as
\ber
{\overline a} & = & {\overline {a}}_{non-res}[1-P_1]\
\eer 
and the total amplitude  for
${ B}^0(v,m) \to D^{*+}(\epsilon_1,v_{+},m_1) D^{*-}(\epsilon_2,v_{-},m_1) 
K_s(p_k)$   
can be written as
\ber
{a} & = & { {a}}_{non-res}[1-P_2]\
\eer 
with
\ber
P_1 & = & \frac{f_{D^{* \prime}_{s1}}}{f_{D^*}}\sqrt{\frac{m^{*\prime}}{m_1}}
\frac{h p_k\cdot v_{-}}{(p_k\cdot v_{-} +m_1-m^{*\prime} +
\frac{i\Gamma_{D^{*\prime}_{s1}}}{2})} \\
P_2 & = & \frac{f_{D^{* \prime}_{s1}}}{f_{D^*}}\sqrt{\frac{m^{*\prime}}{m_1}}
\frac{h p_k\cdot v_{+}}{(p_k\cdot v_{+} +m_1-m^{*\prime} +
\frac{i\Gamma_{D^{*\prime}_{s1}}}{2})} \
\eer
  
In the boosted frame we can write
\ber
p_k\cdot v_{-} & = &\frac{ E_k^{\prime}E_{-}^{\prime} +{p}_k^{\prime} 
{p}_{-}^{\prime}y}{m_{1}} \\
p_k\cdot v_{+} & = &\frac{ E_k^{\prime}E_{+}^{\prime} -{p}_k^{\prime} 
{p}_{+}^{\prime}y}{m_{1}} \
\eer
where $E_k^{\prime}$ and ${p}_k^{\prime}$ are the energy and 
the magnitude of the momentum of the kaon in the boosted frame,    
$E_{\pm}^{\prime}$ and ${p}_{\pm}^{\prime}$ are the energies and 
the magnitude of the momenta of the $D^{*{\pm}}$ in the boosted frame and $m_1$
is the $D^*$ mass.  
In the boosted frame 
we have the following relations
\ber
E_k^{\prime} & = & \gamma(E_k - \vec{\beta}\cdot {\vec{p_k}}) \\
             & = & \frac{1}{\sqrt{1- \frac{E_k^2-m_k^2}{m^2(1-\frac{E_k}{m})^2}}}
             \left[E_k + \frac{E_k^2-m_k^2}{m(1-\frac{E_k}{m})}\right] \\
p_k^{\prime} & = & p_B^{\prime} = \sqrt{ E_k^{\prime 2} -m_k^2} \\             
{p}_{+}^{\prime} & = & {p}_{-}^{\prime}=\sqrt{E_{+}^{\prime 2}-m_1^2} \\
E_{+}^{\prime} & = & E_{-}^{\prime}=\frac{E_B^{\prime}-E_k^{\prime}}{2}\
\eer
where $E_k$ and $p_k$ are the energy and magnitude of the 
momentum of the $K_s$ is the $B$ rest frame, 
 $E_B^{\prime}$ and $p_B^{\prime}$ are the energy and magnitude of the 
momentum
 of the $B$ in the boosted frame and $m$, $m_1$ and $m_k$ are the $B$, $D^*$ and $K_s$ masses.
     
Note from the above relations that $P_1$ and $P_2$ can be expressed in terms of $E_k$ 
and $y$ and  $P_1(y, E_k) =P_2(-y,E_k)$.

Squaring the amplitudes and summing over polarizations one can write
\ber
|{\overline a}|^2 & =&|{\overline a}_{non-res}|^2|1-P_1|^2 \\          
|{ a}|^2 & =&|{ a}_{non-res}|^2|1-P_2|^2 \\          
a^{\ast}{\overline a} & =& {a^{\ast}}_{non-res}{\overline a}_{non-res}
(1-P_2)^{\ast}(1-P_1) \
\eer                    
where
\ber
|{\overline a}_{non-res}|^2 & =&\kappa^2 [-x^2 +2(2x_1x_2 +x_2)x+2x_1^2-
x_2^2+4x_1+2]\\
|{ a}_{non-res}|^2 &=& \kappa^2 [-x^2  +2(2x_1x_2 +x_1)x+2x_2^2-x_1^2+4x_2+
2]\\
 {a^{\ast}}_{non-res}{\overline a}_{non-res} &=& \kappa^2[
 x^2 +(x_1+x_2-2)x+2x_1 +5x_1x_2+2x_2+2 +O(p_k^2/m^2)]\
\eer
where
\bers
\kappa  &=& \frac{G_F}{{\sqrt{2}}} V_c \left(\frac{c_1}{N_c} + c_2 \right) 
 \sqrt{m}\sqrt{m_1}m_1\\
x_1 &=&v \cdot v_{+} = \frac{ E_B^{\prime} E_{+}^{\prime} -
{p}_B^{\prime} {p}_{+}^{\prime}y}{m m_{1}} \\
x_2 &=& v \cdot v_{-}= \frac{ E_B^{\prime} E_{-}^{\prime} +
{p}_B^{\prime} {p}_{-}^{\prime}y}{m m_{1}} \\\\
x & = & v_{+} \cdot v_{-}=\frac{ E_{+}^{\prime} E_{-}^{\prime} +
 {p}_{+}^{\prime}{p}_{-}^{\prime}}
{ m_{1}^2} \\\
\eers
The double differential decay distribution for the time independent
process
$${\bar B}^0(v,m) \to D^{*+}(\epsilon_1,v_{+},m_1) D^{*-}(\epsilon_2,v_{-},m_1) K_s(p_k)$$ 
can be written as
\ber
\frac{1}{\Gamma}\frac{d\Gamma}{dy dE_k} & =& \frac{f(y,E_k)}
{\int f(y,E_k) \frac{p_{k}^{\prime}p_{+}^{\prime}}{m}dydE_k}\\
f(y,E_k) & = &[-x^2 +2(2x_1x_2 +x_2)x+2x_1^2-
x_2^2+4x_1+2]|1-P_1|^2 \
\eer
where $p_{k}^{\prime}$ and $p_{+}^{\prime}$ are 
the magnitudes of the three momentum of the 
$K_s$ and $D^{*+}$ in the boosted frame.

\end{document}